\documentclass[pra, aps, twocolumn]{revtex4}

\usepackage{graphicx}

\usepackage{amssymb, amsmath, color, rotating}
\usepackage{color}

\newcommand{\br}{{\bf r}}
\newcommand{\bk}{{\bf k}}

\newcommand{\bS}{{\bf S}}
\newcommand{\Bm}{{\bf m}}

\begin{document}

\title{Static and Dynamic properties of interacting spin-$1$ bosons in an optical lattice}

\author{Stefan S. Natu}
\email{snatu@umd.edu}
\author{J. H. Pixley}
\author{S. Das Sarma}
\affiliation{Condensed Matter Theory Center and Joint Quantum Institute, Department of Physics, University of Maryland, College Park, Maryland 20742-4111 USA}

\begin{abstract}
We study the physics of interacting spin-$1$ bosons in an optical lattice using a variational Gutzwiller technique. We compute the mean-field ground state wave-function and discuss the evolution of the condensate, spin, nematic, and singlet order parameters across the superfluid-Mott transition. We then extend the Gutzwiller method to derive the equations governing the dynamics of low energy excitations in the lattice. Linearizing these equations, we compute the excitation spectra in the superfluid and Mott phases for both ferromagnetic and antiferromagnetic spin-spin
interactions. In the superfluid phase, we recover the known excitation spectrum obtained from Bogoliubov theory. In the nematic Mott phase, we obtain gapped, quadratically dispersing particle and hole-like collective modes, whereas in the singlet Mott phase, we obtain a non-dispersive gapped mode, corresponding to the breaking of a singlet pair. For the ferromagnetic Mott insulator, the Gutzwiller mean-field theory only yields particle-hole like modes but no Goldstone mode associated with long range spin order. To overcome this limitation, we supplement the Gutzwiller theory with a Schwinger boson mean-field theory which captures super-exchange driven fluctuations. In addition to the gapped particle-hole-like modes, we obtain a gapless quadratically dispersing ferromagnetic spin-wave Goldstone mode. We discuss the evolution of the singlet gap, particle-hole gap, and the effective mass of the ferromagnetic Goldstone mode as the superfluid-Mott phase boundary is approached from the insulating side. We discuss the relevance and validity of Gutzwiller mean-field theories to spinful systems, and potential extensions of this framework to include more exotic physics which appears in the presence of spin-orbit coupling or artificial gauge fields. 
\end{abstract}
\maketitle

\section{Introduction}

Recent progress in ultra-cold atoms has made it possible to study strongly correlated phenomena in a number of different contexts, which have no natural analog in condensed matter physics. One such example is spinless and spinful many-body \textit{bosonic} systems, whose rich physics has been experimentally explored in the continuum and in optical lattices \cite{greiner, jaksch, jason, machida, stenger, stamper-kurn}. Large spin systems offer interesting possibilities to study the interplay between competing/complimentary orders at zero and finite temperatures such as single-particle and pair superfluidity, spin and liquid crystallinity, all of which can be probed using a variety of experimental tools \cite{muellerrotating, dienerho, natuspin1, imambekov}. Recently, attention has turned to the physics of bosonic and fermionic systems in the presence of spin-orbit coupling or artificial gauge fields \cite{nistexpt, spielman2, chinasoc, ji, zhang1, zwierleinsoc, miyakebutterfly, blochbutterfly}. Spin-orbit coupling and artificial gauge fields introduce degeneracies in the single-particle spectrum, which tends to frustrate the usual Bose condensation, setting the stage for the appearance of exotic ordered phases even at the mean-field level, such as striped Bose condensates which spontaneously break translational symmetry \cite{jasonspinorbit, stringari, wang, stanescu} or magnetized spin stripe phases which spontaneously break time-reversal symmetry \cite{congjun}. Furthermore, single particle degeneracies can amplify the role of quantum fluctuations leading to  chiral superfluids \cite{xiaopeng} or bosonic phases with topological order, which resemble the integer and fractional quantum Hall effect \cite{cooper} of electrons in a magnetic field. 

With these exciting experimental advances, it has become increasingly important to develop theoretical methods which are sophisticated enough to describe the multitude of order parameters and broken symmetry phases that can potentially occur in these interesting interacting systems, even at the mean-field level. 
In addition, it is essential to first gain significant insight into the various forms of order such systems can develop before introducing the next level of complexity through artificial gauge fields or spin orbit coupling.
With this in mind, we perform a variational Gutzwiller study of the mean-field physics of a spin-$1$ Bose gas in an optical lattice, in the absence of spin-orbit coupling. We highlight the key virtues and limitations of this method in describing spinful bosonic systems, by giving a comprehensive account of the static and dynamic properties of the spin-$1$ Bose gas in an optical lattice. We then supplement this method by a Schwinger boson mean-field theory to capture the gapless Goldstone mode in the ferromagnetic Mott phase. 
We discuss extensions of this approach to the spin-orbit coupled, large spin problem.

The bosonic Gutzwiller mean-field technique, introduced by Rokhsar and Kotliar \cite{rokhsar}, describes the mean-field physics of the Bose Hubbard model \cite{fisher}, which is realized by trapping bosons in a deep optical lattice \cite{jaksch, greiner}. Known to be exact in infinite dimensions, the technique captures local physics by decomposing the full Bose Hubbard Hamiltonian into a sum of on-site Hamiltonians coupled by mean-fields. The transition from the superfluid to the Mott insulator is then obtained by self-consistently solving for where the mean-field order parameter vanishes. Within the Mott lobes, the Gutzwiller Hamiltonian is therefore purely local (\textit{i.e.}, identical to the zero hopping limit), and hence does not \textit{a priori} capture any correlations. In the spinless case, where exact Quantum Monte Carlo (QMC) simulations are possible \cite{prokofiev}, the Gutzwiller method is known to correctly capture the qualitative features of the phase diagram. We will apply the Gutzwiller technique to spin-$1$ bosons in the current work.

The equilibrium physics of this model for spinless bosons is well known, and recently it has been extended to include spinor systems, such as the one we study here \cite{kimura, pai, krutitsky, demler, scalettar, batrouni}. For spinless bosons, this technique has recently also been extended to finite clusters, where the physics within an $m \times n$ plaquette is solved exactly, and the plaquettes are coupled by mean-fields. Systematic studies using this cluster Gutzwiller method have shown that for relatively small cluster sizes, \textit{quantitative} agreement is obtained for the location of the phase boundaries with numerically exact Quantum Monte Carlo studies \cite{luhmann}. Importantly, such cluster methods offer a comparatively numerically efficient way to systematically incorporate correlation effects and perform non-equilibrium dynamics, which is highly relevant to ongoing experiments on strongly correlated bosons \cite{chengslow, greinerfast, demarco1, demarco2}.

In this paper we provide a systematic study of the physics of the spin-$1$ Bose Hubbard model.
We discuss the equilibrium theory and highlight the advantages and disadvantages of the Gutzwiller approach in correctly capturing the spin physics in an interacting bosonic system. Although the equilibrium phase boundaries for this model have been established by several authors \cite{kimura, pai, krutitsky}, a systematic study of the various order parameters present in the spin-$1$ Bose Hubbard model has been lacking.  As this is essential in forming a comprehensive understanding of each phase, a new feature of our work is a focus on the evolution of the various \textit{order parameters}, such as the spin, nematic and singlet order. As these order parameters can be directly
measured in experiments, such a study is a necessity in the interpretation of the experiment and the understanding of the spinful Bose-Hubbard quantum phase diagram.

Recently, several authors \cite{navez, menotti, natudynamics} have extended the Gutzwiller technique to capture \textit{dynamics}, by linearizing the mean-field equations of motion about the superfluid and Mott insulating ground states. Indeed the linearized theory correctly captures the well-known Bogoliubov modes in the superfluid, the gapped particle-hole like modes in the Mott insulator, and correctly describes qualitative aspects of how these modes evolve across the superfluid-Mott phase boundary. Here we extend the time-dependent Gutzwiller framework developed by Krutitsky and Navez \cite{navez} to spin-$1$ Bose Hubbard model, and compute the low energy spectrum across the entire superfluid-Mott phase diagram for ferromagnetic and anti-ferromagnetic spin-dependent interactions \cite{shinozaki}. In particular, we show that the Gutzwiller approach by itself, fails to fully capture the spin physics in the Mott insulator, and therefore we supplement this theory with a Schwinger boson mean-field theory, which captures inter-site magnetic fluctuations. Our combined Gutzwiller + Schwinger boson approach thus more or less fully captures the low energy mean-field properties of the spin-$1$ Bose Hubbard model, and sets the stage for studies of more complicated correlated bosonic Hamiltonians, which include spin-orbit coupling or artificial gauge fields, which we leave for future study. 

The rest of this paper is organized as follows: in Sec. II, we present the Gutzwiller mean-field equations, and define the various order-parameters which we use to compute the ground state properties and the superfluid-insulator phase boundaries. In Sec. III, we present the Gutzwiller phase diagrams for ferromagnetic and anti-ferromagnetic interactions focussing on the evolution of the spin, nematic and singlet order parameters across the entire phase diagram, which are absent in the spinless case. In Sec. IV, we linearize around the Gutzwiller ground state   to produce the equations for the low energy collective modes. In Sec. V, we 
present the excitation spectra for ferromagnetic and anti-ferromagnetic interactions in the superfluid and Mott phases.
In Sec. VI, we discuss the relevance of our results to experiments and in Sec. VII, we provide a summary of our results and discuss directions for future work.
Readers uninterested in the detailed theoretical derivations of the equations may skip Sections  II and IV, which are technical in nature. 

\section{Gutzwiller mean-field theory: Statics}

In this section we outline the Gutzwiller mean-field theory which we use to compute the ground state energy, wave-function and order parameters in the following sections. We remark that the Gutzwiller mean-field equations have been obtained previously by several authors \cite{pai, kimura, krutitsky} and are only reproduced here for completeness (and for providing a background, as well as a context for the new results obtained by us for the spin-1 bosons). Our starting point is the generalized Hamiltonian of the spin-$1$ Bose gas in an optical lattice \cite{imambekov}
\begin{eqnarray}\label{ham}
H-\mu \hat{N} &=& -t\sum_{\langle i,j \rangle,\alpha}\left(a_{i\alpha}^{\dag}a_{j\alpha}+\mathrm{H.c} \right)+ \frac{U_0}{2}\sum_i \hat{n}_i(\hat{n}_i-1)
\nonumber
\\
&+& \frac{U_2}{2}\sum_i \left(\bS^2-2\hat{n}_i \right)-\mu\sum_i \hat{n}_i  
\end{eqnarray}
where $\langle i,j \rangle$ denotes nearest neighbor sites, a chemical potential $\mu$, and we have defined $\hat{n}_i = \sum_{\alpha}a_{i\alpha}^{\dag}a_{i\alpha}$, $\bS=\sum_{\alpha,\beta}a_{i\alpha}^{\dag}{\bf T}_{\alpha\beta}a_{i\beta}$, where the ${\bf T}$ are a vector of spin-1 matrices.
While the mean field equations are generic for any dimensionality and lattice geometry, in the following we focus on a three dimensional cubic lattice.

To the derive the Gutzwiller equations, we proceed by writing the tight binding model at the mean field level by treating each neighbor of site $i$ using the mean field level truncation
\begin{equation}
a_{i\alpha}^{\dag}a_{j\alpha} \rightarrow \langle a_{i\alpha}^{\dag} \rangle a_{j\alpha} +  a_{i\alpha}^{\dag}  \langle a_{j\alpha} \rangle.
\end{equation}
\\
Following the standard Gutzwiller mean-field theory, we write the ground state wave function in the Fock basis as a direct product over single site wave functions as
\begin{eqnarray}\label{gutzgs}
|\Psi_{GS}\rangle &=& \otimes_{i=1}^{N_{\mathrm{site}}}|\phi_i\rangle
\\
|\phi_i\rangle &=& \sum_{m_{-1},m_0,m_1}A_{m_{-1}m_0m_1}|m_{-1},m_0,m_1\rangle
\end{eqnarray}
where $m_{-1},m_0,m_1$ denote Fock states in the $m_z=-1,0,1$ state respectively, and we have assumed the $A_{m_{-1}m_0m_1}$ to be site independent.  We will now evaluate the Hamiltonian (Eq.~\ref{ham}) with respect to the mean-field ansatz (Eq.~\ref{gutzgs}).  The terms diagonal in particle number yield
\begin{eqnarray}
& &\langle \phi_i | \frac{U_0}{2}\hat{n}_i(\hat{n}_i-1) -\mu\hat{n}_i|\phi_i \rangle =
\sum_{m_{-1},m_0,m_1}|A_{m_{-1}m_0m_1}|^2 
\nonumber
\\
\nonumber
& &\times\Bigg(\frac{U_0}{2}(m_{-1}+m_0+m_1)(m_{-1}+m_0+m_1-1) 
\\
& &-\mu(m_{-1}+m_0+m_1) 
\Bigg).
\nonumber
\end{eqnarray}
The spin operator $\bS^2$ yields off diagonal terms in Fock space. In particular we obtain
\begin{widetext}
\begin{eqnarray}\label{u2eq}
\langle \phi_i | \frac{U_2}{2}(\bS_i^2-2\hat{n}_i )|\phi_i \rangle &=& U_2 \sum_{m_{-1},m_0,m_1}\Bigg[|A_{m_{-1}m_0m_1}|^2\Big( \frac{1}{2}[(m_1 - m_{-1})^2-m_1-m_{-1}]+m_1m_0+m_{-1}m_0 \Big)
\\\nonumber
&+& A^*_{m_{-1}m_0m_1}A_{(m_{-1}-1)(m_0+2)(m_1-1)}\sqrt{m_1(m_0+1)(m_0+2)m_{-1}}
\\\nonumber
&+&
A^*_{m_{-1}m_0m_1}A_{(m_{-1}+1)(m_0-2)(m_1+1)}\sqrt{(m_1+1)m_0(m_0-1)(m_{-1}+1)}
\Bigg].
\end{eqnarray}
\end{widetext}
Now that we have all of the two particle terms we can move on to calculating the expectation value of hopping (\textit{i.e.} $t$) dependent terms.  For the creation operator this yields
\begin{eqnarray}
\langle \phi_i | -t\sum_{\alpha} a_{i\alpha}^{\dag}|\phi_i \rangle= -2zt\sum_{m_{-1},m_0,m_1}A^*_{m_{-1}m_0m_1}\\\nonumber\Big[A_{m_{-1}m_0(m_1-1)}\sqrt{m_{1}}+A_{m_{-1}(m_0-1)m_1}\sqrt{m_{0}}
\\
\nonumber+
A_{(m_{-1}-1)m_0m_1}\sqrt{m_{-1}}\Big],
\end{eqnarray}
where $z$ denotes the number of nearest neighbors, and for the destruction operator this yields
\begin{eqnarray}
\langle \phi_i | -t\sum_{\alpha}a_{i\alpha} |\phi_i \rangle = -2zt\sum_{m_{-1},m_0,m_1}A^*_{m_{-1}m_0m_1}\\\nonumber\Big[A_{(m_{-1}+1)m_0m_1}\sqrt{m_{-1}+1}+
A_{m_{-1}(m_0+1)m_1}\sqrt{m_{0}+1} 
\\
\nonumber
+A_{m_{-1}m_0(m_1+1)}\sqrt{m_{1}+1}\Big].
\end{eqnarray}
We self consistently determine the values of $\langle a_{j\beta} \rangle$, by viewing the ground state expectation value of the Hamiltonian as ${\cal{A}} .\mathcal{H}.{\cal{A}}$, where ${\cal{A}}$ is a vector of all the $A_{m_{-1}m_0m_1}$, and we just have to diagonalize $\mathcal{H}=\langle n_1,n_0,n_{-1}|H|m_{-1},m_0,m_1 \rangle$ iteratively.  This is equivalent to determining the $A_{m_{-1}m_0m_1}$ coefficients variationally through solving
\begin{equation}
 \frac{\delta}{\delta A_{m_{-1}m_0m_1}^*} \langle \Psi_{GS} | H - \mu N | \Psi_{GS} \rangle =0,
\end{equation}
and similarly for $A_{m_{-1}m_0m_1}$. At each step of the numerical iteration, we self-consistently determine the mean-field $\langle a_{i\alpha} \rangle$, which is identical on every site in the homogeneous system we study.  In what follows, we therefore occasionally drop the site index when unnecessary. 

We note that the mean-field theory described here can be readily generalized to capture translational symmetry breaking phases by allowing the mean-field to vary on every site. However, absent spin-orbit coupling or artificial gauge fields, we do not expect translational symmetry breaking phases to occur, and therefore restrict our study to spatially homogeneous mean fields.

\subsection{Order parameters}

The spin-$1$ system generally possesses four order parameters namely, the condensate fraction, the spin, the nematic director, and the singlet order parameter, which we determine using the ground state wave-function computed above. The condensate fraction in the three hyperfine spin states can be expressed as $\langle a \rangle = \{\langle a_{-1} \rangle,\langle a_{0} \rangle,\langle a_{1} \rangle\} $ and is given simply by
\begin{eqnarray}\label{condfrac}
\langle a \rangle = \sum_{m_{-1}m_{0},m_{1}}\Bigg(A_{m_{-1}m_{1}m_{0}}
\Big\{A^{*}_{(m_{-1}-1)m_{0}m_{1}}\sqrt{m_{-1}},
\nonumber
\\
A^{*}_{m_{-1}(m_{0}-1)m_{1}}\sqrt{m_{0}},A^{*}_{m_{-1}m_{0}(m_{1}-1)}\sqrt{m_{1}}\Big\}\Bigg).
\end{eqnarray}

The spin is a single-particle vector order parameter given by $ \langle \bS \rangle$ in the mean-field ground state. The nematic director is a two-particle tensor order parameter which is  given by the matrix $\mathcal N_{\alpha\beta} = \frac{1}{2}\langle S_{\alpha}S_{\beta} + S_{\beta}S_{\alpha} \rangle$, where $\alpha,\beta \in \{x, y, z\}$ correspond to the spin-operators representing the $x$, $y$, and $z$, components of the spin. Diagonalizing the nematic tensor yields three eigenvalues, the largest of which (denoted by $\lambda_{\cal{N}}$) corresponds to the degree of nematicity. The singlet order parameter is also a two-particle order parameter which measures the number of singlets on a given site. It is a scalar object obtained by computing the number of singlets $\langle\Theta^{\dagger}_{i}\Theta_{i}\rangle$, in the mean-field ground state, where $\Theta^{\dagger}_{i} = 2a^{\dagger}_{1i}a^{\dagger}_{-1i} -a^{\dagger}_{0i}a^{\dagger}_{0i}$ (see Refs. \onlinecite{hofrag, law}) is the singlet creation operator on site $i$. Once the ground state wave-function is obtained, each of these order parameters can be readily computed numerically.

\section{Phase Diagram}

\begin{figure}
\begin{picture}(200, 180)
\put(-15, -10){\includegraphics[scale=0.5]{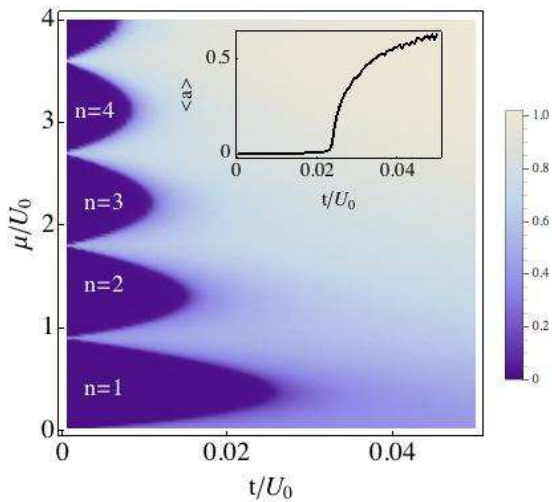}}
\end{picture}
\caption{\label{ferrofig} (Color Online) Density plot showing the evolution of the superfluid order parameter in the ferromagnetic ($U_{2} =-0.1U_0$) spin-$1$ Bose gas. The superfluid order parameter goes to zero at the superfluid-Mott phase boundary as shown in the inset. The total spin $\mathcal{S}/n = 1$ throughout the phase diagram. The numbers label the density in the Mott lobes. All transitions here are second order.}
\end{figure}

In this section we present the phase diagram of the spin-$1$ Bose Hubbard model and highlight the key virtues and shortcomings of the Gutzwiller approach with respect to other approaches, namely the weak coupling Bogoliubov theory of the spin-$1$ Bose gas~\cite{machida, jason} and the strong coupling expansion \cite{imambekov, snoek}. We stress here that while our results for the phase boundaries are not new, and have been discussed by several authors \cite{pai, kimura, krutitsky, scalettar, batrouni}, we focus on the evolution of the spin, nematic, and singlet order parameters across the superfluid-Mott transition, which has not yet been discussed in the literature and is essential to understand and characterize the nature of each phase as manifested in experiments. Importantly, these order parameters distinguish the spin-$1$ gas from its well studied spinless counterpart and are generically present in higher spin systems, such as spin-$3$ Chromium atoms \cite{dienerho}. Understanding how these evolve in the spin-$1$ Bose Hubbard model is therefore crucial to developing theories of larger spin systems, which are currently being explored experimentally \cite{pfau, erbium, dysprosium}.

\subsection{Ferromagnetic interactions}

We start by discussing the conceptually simple ferromagnetic case for $U_{2} <0$, which occurs for $^{87}$Rb, where the system has only two order parameters, the superfluid order parameter and the total spin $\mathcal{S}=\sqrt{\langle S_{x} \rangle^{2} + \langle S_{y} \rangle^{2} + \langle S_{z} \rangle^{2}}$.

\begin{figure*}
\begin{picture}(200, 180)
\put(-127, 170){\text{(a)}}
\put(100, 170){\text{(b)}}
\put(-115, -10){\includegraphics[scale=0.5]{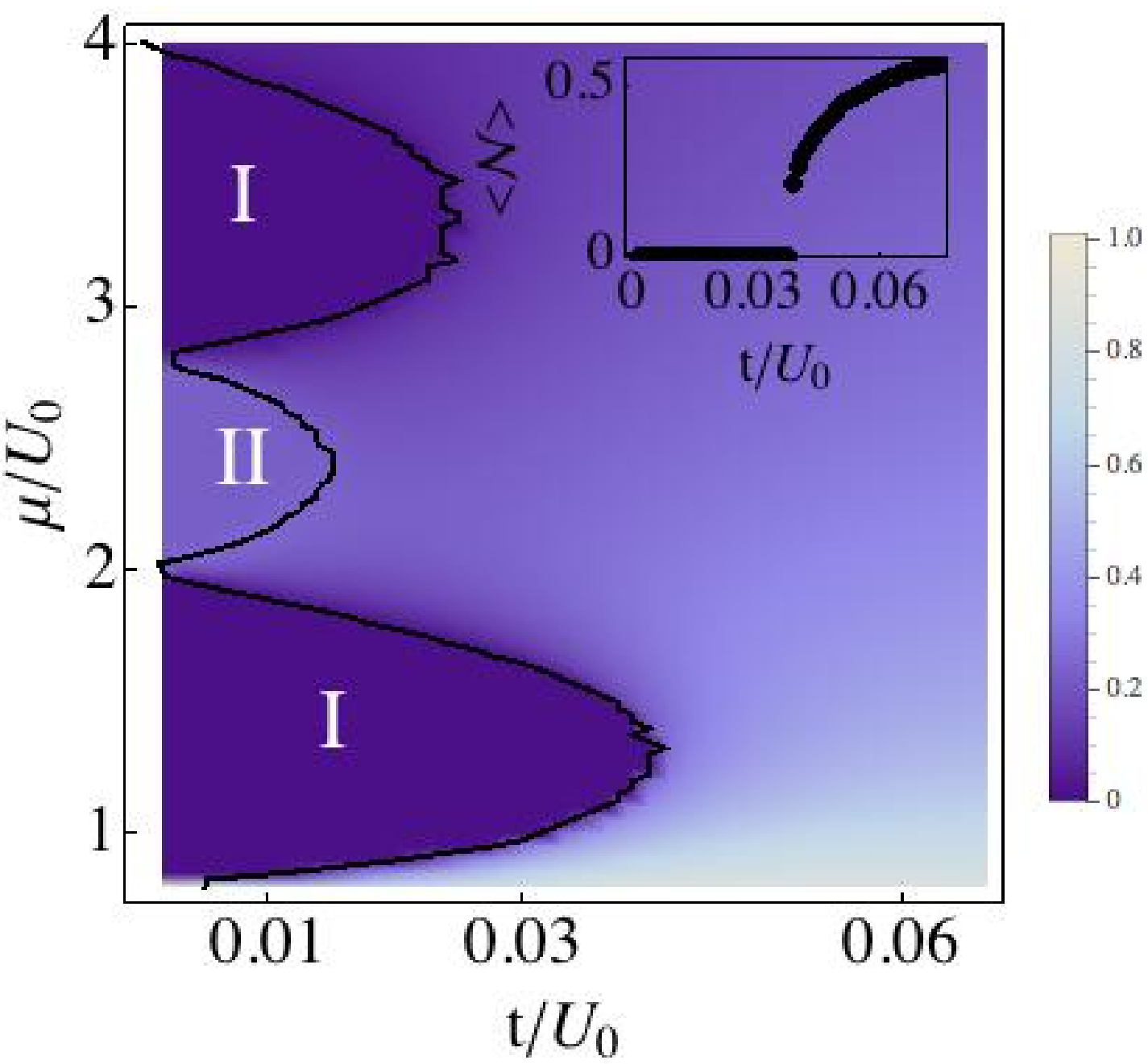}}
\put(115, -10){\includegraphics[scale=0.5]{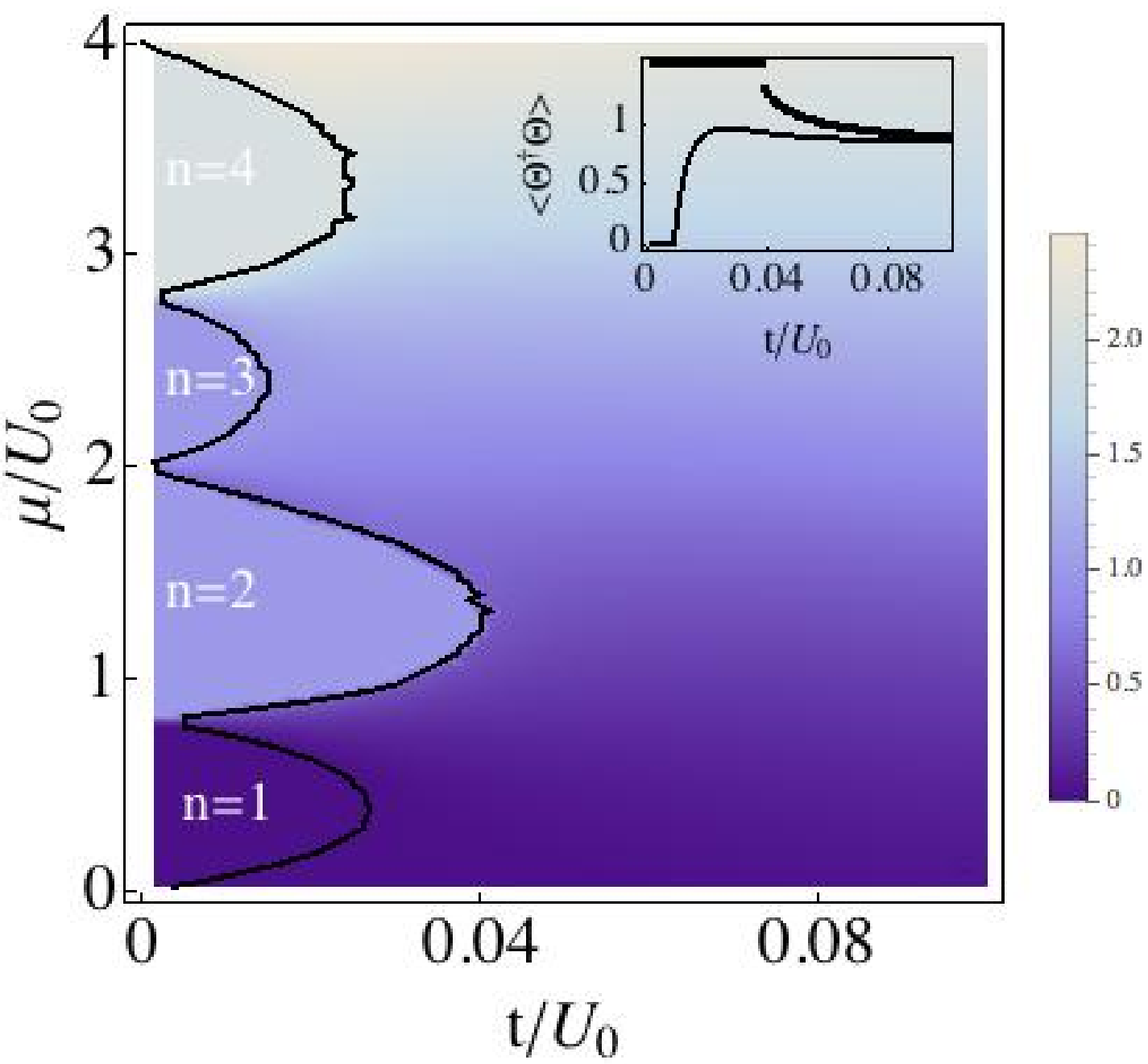}}
\end{picture}
\caption{\label{antiferrofig} (Color Online) Density plot showing the evolution of the nematic (left) and singlet (right) order parameters in the anti-ferromagnetic ($U_{2} =0.1U_0$) spin-$1$ Bose gas. The nematic order parameter is computed by taking the maximum eigenvalue of the nematic tensor. In the singlet phases, labelled $I$, nematic order vanishes, and the superfluid-Mott transition is first order, as indicated by the finite jump in the nematic order, as shown in the inset. In contrast, the transition into the lobe labelled $II$ is a continuous second order quantum phase transition. On the right, the singlet number takes on an integer value equal to $1(2)$ in the $n=2(4)$ Mott lobes, corresponding to the number of local singlets per site. The inset shows the singlet order parameter for two different cuts across the phase diagram: the cuts traverse the $n=2$ (dashed) and $n=1$ (solid) Mott-superfluid phase boundaries. At the $n=2$ Mott-superfluid phase boundary, the singlet order parameter displays a finite jump, which is evidence of the first order nature of the transition.}
\end{figure*}

In Fig.~\ref{ferrofig}, we plot the theoretically calculated evolution of the superfluid order parameter, which clearly shows the superfluid-Mott phase boundary, where the superfluid order parameter vanishes. 
The second order nature of the phase transition \cite{krutitsky, pai, kimura} is clear from the smooth manner in which the order parameter goes to zero.
This is to be expected, as on the ferromagnetic side, the total spin simply locks to the density, and the system resembles a spinless gas, which is known to have a second order superfluid-Mott insulator transition. The total spin 
$\mathcal{S}/n$
normalized by the total density is equal to unity throughout the phase diagram. 

A curious feature of this mean-field theory is that the dependence on $U_{2}$ completely drops out for the $n=1$ Mott lobe, as is evident from Eq.~\ref{u2eq}. This is because, absent the mean-field term, number fluctuations are completely frozen out and the $n=1$ Mott lobe is obtained by setting either $m_{1}$, $m_{0}$ or $m_{-1} =1$, while the others are zero. The physics of the $n=1$ Mott insulator thus has to be inferred by continuity from the superfluid side. For $U_{2} <0$, where the superfluid is ferromagnetic, the Mott insulator is also ferromagnetic, whereas for $U_{2} > 0$, where the superfluid is polar, the Mott insulator is 
nematic. A better treatment of the ferromagnetic Mott insulator, which captures spin fluctuations can be done using a Schwinger boson mean-field theory, described in detail in Sec. VB.1.

\subsection{Anti-ferromagnetic interactions}

We now turn to the more interesting case of anti-ferromagnetic interactions, which are present in $^{23}$Na. Here the system is described by three order parameters, the complex superfluid order parameter, the tensor nematic order parameter ${\cal{N}}/n^{2}$, and the scalar singlet order parameter, which measures the number of singlets created on a given site. 

In Fig.~\ref{antiferrofig}, we present the phase diagram for anti-ferromagnetic spin-dependent interactions $U_{2}> 0$ in terms of the nematic 
 and the singlet order parameters.
 To display the nematic order parameter, we compute the largest eigenvalue $\lambda_{\cal{N}}$ of the nematic tensor ${\cal{N}}/n^{2}$ for every value of $t/U_{0}$ and $\mu/U_{0}$ throughout the phase diagram. The inset shows a cut of the nematic order parameter at fixed $\mu$ through the $n=2$ Mott lobe. The labels $I$ and $II$ indicate the nature of the superfluid-Mott insulator transition as being first and second order respectively.

In the superfluid phase, we recover the well known continuum result, namely that the largest eigenvalue of the nematic tensor is $\lambda_{\cal{N}} \rightarrow 0.5$. In the superfluid, the spin, density and nematic orders compete \cite{muellerrotating}, which enforces a constraint on the nematic tensor such that the sum of the eigenvalues equals $1$. In the even integer Mott lobes, the system enters the singlet phase, which is characterized by zero total spin $\langle \textbf{S} \rangle = 0$, and zero nematicity. In other words, all three eigenvalues of the nematic tensor are zero in this phase. Unlike the ferromagnetic gas, several authors have argued that the transition from the singlet Mott insulator to the nematic superfluid is \textit{first order} \cite{kimura, pai, krutitsky}, and this has been confirmed numerically in $1$ and $2$D through exact Quantum Monte Carlo simulations \cite{scalettar, batrouni}.

The first order nature of the transition should also be evident in the evolution of the nematic, condensate and singlet order parameter across the phase boundary. In the inset on the left panel, we plot a cut through the nematic order parameter across the $n=2$ Mott lobe, which clearly shows a discrete jump at the superfluid-singlet Mott transition. As the nematic order can be probed experimentally using optical birefringence techniques \cite{carusotto}, this is an example of a first order quantum phase transition that can be readily studied in the laboratory. Additionally, the superfluid order parameter also shows a discrete jump at this transition to zero, unlike for a second order transition. Near the odd integer Mott lobes, the transition is once again second order. 

Unlike the nematic order parameter, which is identically zero in the singlet Mott insulator, the singlet order parameter $\langle \Theta^{\dagger}\Theta \rangle$ takes on an integer value equal to half the total particle number in the even Mott lobes. The inset on the right panel shows two horizontal cuts through the phase diagram across the superfluid - $n$ Mott transition where $n=1$ (solid) and $n=2$ (dashed). The singlet order parameter is zero in the $n=1$ Mott lobe, as expected, and precisely $1$ in the $n=2$ Mott lobe. Furthermore, the singlet order parameter shows a small but finite jump at the Mott-superfluid transition, once again revealing the first order nature of the transition. 

A first order superfluid-Mott transition usually implies a small but finite coexistence region where a metastable superfluid phase can occur in addition to a Mott insulator with local singlets. Such a coexistence region has been discussed within the mean-field context \cite{krutitsky} and validated through exact Quantum Monte Carlo simulations in $1$D \cite{scalettar}. Nonetheless, the first order transition can be readily probed by studying the evolution of the superfluid, nematic or singlet order parameters. Precisely figuring out whether this transition is indeed first order or an artifact of the mean-field approximation will demand more sophisticated numerical simulations (e.q. QMC) in two and three dimensions, which our work should motivate. 

For very small $U_{2}/U_{0}$, Imambekov \textit{et al.} \cite{imambekov} find an additional first order phase transition within the Mott lobe, which corresponds to a transition from a nematic Mott phase to a singlet phase. This transition is not captured by the Gutzwiller mean-field theory, because it does not include any spin fluctuations in the Mott insulator. Nonetheless, the Gutzwiller theory gets the correct local spin physics at zero hopping, and extrapolates this wave-function throughout the entire Mott lobe. Note that within this mean field theory, the singlet Mott lobes are larger than the nematic lobes. This is because the tendency to form local singlets which is favored by the repulsive spin-dependent interactions tends to destroy superfluid order more easily, thus enhancing the Mott region of the phase diagram. 

\section{Gutzwiller mean-field theory: Dynamics}

Having formed a more comprehensive understanding of
 the static ground state quantum phase diagram using the Gutzwiller technique, we now turn our attention to the low lying excitation spectrum in the spin-$1$ gas. Some of the results derived here for $U_{2}>0$ were recently obtained in Ref.~\onlinecite{shinozaki}. Here we present some more details on the excitation spectrum for $U_{2} >0$, and also describe the excitations on the ferromagnetic side $U_{2} <0$. We start by presenting a complete derivation of the equations of motion, which describe the full mean-field dynamics about the Gutzwiller ground state, generalizing the earlier work in Ref.~\cite{navez} for the spinless Bose Hubbard model. 

To begin, we generalize the Gutzwiller wave-function to include dynamics
\begin{equation}
|\phi_i\rangle = \sum_{m_{-1},m_0,m_1}A_{m_{-1}m_0m_1}(i,t)|m_{-1},m_0,m_1\rangle
\end{equation}
and the coefficients explicitly depend on the site index $i$ and the time $t$. We now wish to variationally minimize the time dependent Schr\"odinger equation with respect to $A_{m_{-1}m_0m_1}$, we have
\begin{equation}
 \frac{\delta}{\delta A_{m_{-1}m_0m_1}^*} \langle \Psi_{GS} |i\partial_t- H + \mu N | \Psi_{GS} \rangle =0.
\end{equation}
Taking the variational derivative, we arrive at
\begin{eqnarray}\label{aeq}
& &i\partial_t A_{\Bm} = D_{\Bm}A_{\Bm}+S_{\Bm}^{+-}A_{(m_{-1}-1)(m_0+2)(m_1-1)} 
\\
\nonumber 
&+& S_{\Bm}^{-+}A_{(m_{-1}+1)(m_{0}-2)(m_1+1)} -t\Big[\sqrt{m_{-1}} A_{(m_{-1}-1)m_0m_1}
\\
\nonumber 
&+& \sqrt{m_{0}}A_{m_{-1}(m_0-1)m_1}+\sqrt{m_{1}}A_{m_{-1}m_0(m_1-1)}\Big]
\\
\nonumber
& -&  t\Big[\sqrt{m_{-1}+1}A_{(m_{-1}+1)m_0m_1}+\sqrt{m_{0}+1}A_{m_{-1}(m_0+1)m_1}
 \\
 \nonumber
 &+&\sqrt{m_{1}+1}A_{m_{-1}m_0(m_1+1)}\Big],
\end{eqnarray}
where we have introduced the shorthand notation $\Bm = (m_{-1},m_0,m_1)$, and the dependence of $A_{\Bm}$ on space and time $(i, t)$ is implicit. We have also defined the following $m$ dependent terms to simplify the writing
\begin{eqnarray}
 D_{\Bm} &=& \frac{U_0}{2}(m_{-1}+m_0+m_1)(m_{-1}+m_0+  m_1 -1) 
 \nonumber
 \\
 \nonumber
&+&U_2\Big( \frac{1}{2}[(m_1 - m_{-1})^2-m_1-m_{-1}]+ m_1m_0
\\
&+&m_{-1}m_0 \Big)-\mu(m_{-1}+m_0+m_1) 
\\
\nonumber
S_{\Bm}^{+-} &=& U_2\sqrt{m_1(m_0+1)(m_0+2)m_{-1}},
\\
\nonumber
S_{\Bm}^{-+} &=& U_2\sqrt{(m_1+1)m_0(m_0-1)(m_{-1}+1)}.
\end{eqnarray}

To obtain the low energy spectrum, we follow Ref.~\onlinecite{krutitsky}, and expand the wave-function coefficients about the mean-field solution
\begin{equation}
A_{\Bm}(i,t) \approx e^{-E_0 t}(A^{(0)}_{\Bm}+A^{(1)}_{\Bm}(i,t)),
\end{equation}
where $E_0$ is the ground state energy, $A_{\Bm}^{(0)}$ is the mean field solution and $A_{\Bm}^{(1)}$ are the fluctuations.
We then expand the space and time dependence of $A_{\Bm}^{(1)}$ into plane wave states to obtain
\begin{equation}\label{wfexpand}
A^{(1)}_{\Bm}(i,t) = u_{\bk,\Bm}e^{i(\bk\cdot \br_i-i\omega_{\bk}t)} + v_{\bk,\Bm}^*e^{-i(\bk\cdot \br_i-i\omega_{\bk}t)},
\end{equation}
where $\omega_{\textbf{k}}$ is the low energy dispersion we are trying to calculate. Inserting this ansatz into Eq.~\ref{aeq}, and keeping only terms linear in $u_{\textbf{k}}$ and $v_{\textbf{k}}$, we obtain
\begin{widetext}
\begin{eqnarray}
\omega_{\bk}u_{\bk,\Bm} &=& (D_{\Bm}-E_0)u_{\bk,\Bm} + S_{\Bm}^{+-}u_{\bk,(m_{-1}-1)(m_0+2)(m_1-1)}+ S_{\Bm}^{-+}u_{\bk,(m_{-1}+1)(m_0-2)(m_1+1)}
\\
\nonumber
&-&t\left(\sqrt{m_{-1}}\psi^{(0)}_{m_{-1}}u_{\bk,(m_{-1}-1)m_0m_1} +\sqrt{m_{0}}\psi^{(0)}_{m_0}u_{\bk,m_{-1}(m_0-1)m_1}+\sqrt{m_{1}}\psi^{(0)}_{m_1}u_{\bk,m_{-1}m_0(m_1-1)} \right)
\\
\nonumber
&-&t\left(\sqrt{m_{-1}+1}\psi^{(0)*}_{m_{-1}}u_{\bk,(m_{-1}+1)m_0m_1} +\sqrt{m_{0}+1}\psi^{(0)*}_{m_0}u_{\bk,m_{-1}(m_0+1)m_1}+\sqrt{m_{1}+1}\psi^{(0)*}_{m_1}u_{\bk,m_{-1}m_0(m_1+1)} \right)
\\
\nonumber
&-&t\gamma_{\bk}\left(\sqrt{m_{-1}}U^+_{m_{-1}}A^{(0)}_{(m_{-1}-1)m_0m_1} + \sqrt{m_{0}}U^+_{m_{0}}A^{(0)}_{m_{-1}(m_0-1)m_1}
+\sqrt{m_{1}}U^+_{m_{1}}A^{(0)}_{m_{-1}m_0(m_1-1)}\right)
\\
\nonumber
&-&t\gamma_{\bk} \left(\sqrt{m_{-1}+1}U^{-}_{m_{-1}}A^{(0)}_{(m_{-1}+1)m_0m_1} + \sqrt{m_{0}+1}U^{-}_{m_{0}}A^{(0)}_{m_{-1}(m_0+1)m_1} + \sqrt{m_{1}+1}U^{-}_{m_{1}}A^{(0)}_{m_{-1}m_0(m_1+1)}\right).
\end{eqnarray}
\end{widetext}
where we have introduced $\psi^{(0)}_{\alpha} = 2 z\langle a_{\alpha} \rangle$, $\psi^{(0)*}_{\alpha} = 2 z\langle a_{\alpha}^{\dag} \rangle$ [the $(0)$ here denotes an evaluation with respect to the mean field ground state $A_{\Bm}^{(0)}$], 
\begin{equation}
\gamma_{\bk} = 2\left(\cos(k_x)+\cos(k_y)+\cos(k_z) \right)
\label{eqn:gammak}
\end{equation}
and
\begin{eqnarray}
U^+_{m_{\alpha}} &=& \sum_{\Bm'}\sqrt{m_{\alpha}'+1}\hspace{20mm}
\\
\nonumber
&\times&\left(A_{\Bm'}^{*(0)}u_{\bk,(m_{\alpha}'+1)m_{\gamma}'m_{\delta}' } +
A_{(m_{\alpha}'+1)m_{\gamma}'m_{\delta}' }^{(0)}v_{\bk,\Bm'}\right)
\\
U^-_{m_{\alpha}} &=& \sum_{\Bm'}\sqrt{m_{\alpha}'}\hspace{20mm}
\\
\nonumber
&\times&\left(A_{\Bm'}^{*(0)}u_{\bk,(m_{\alpha}'-1)m_{\gamma}'m_{\delta}' }+A_{(m_{\alpha}'-1)m_{\gamma}'m_{\delta}' }^{(0)}v_{\bk,\Bm'}\right)
\end{eqnarray}
where $\alpha\neq\gamma\neq\delta \in \{-1, 0, 1\}$. For example for $\alpha=0$ this would yield
\begin{eqnarray}
U^+_{m_0} &=& \sum_{\Bm'}\sqrt{m_{0}'+1}\hspace{20mm}
\nonumber
\\
\nonumber
&\times&\left(A_{\Bm'}^{*(0)}u_{\bk,m_{-1}'(m_{0}'+1)m_{1}' }+A_{m_{-1}'(m_{0}'+1)m_{1}' }^{(0)}v_{\bk,\Bm'}\right).
\end{eqnarray}
Similarly, for the $v$s, we obtain
\begin{widetext}
\begin{eqnarray}
-\omega_{\bk}v^*_{\bk,\Bm} &=& (D_{\Bm}-E_0)v^*_{\bk,\Bm} + S_{\Bm}^{+-}v^*_{\bk,(m_{-1}-1)(m_0+2)(m_1-1)}+ S_{\Bm}^{-+}v^*_{\bk,(m_{-1}+1)(m_0-2)(m_1+1)}
\\
\nonumber
&-&t\left(\sqrt{m_{-1}}\psi^{(0)}_{m_{-1}}v^*_{\bk,(m_{-1}-1)m_0m_1} +\sqrt{m_{0}}\psi^{(0)}_{m_{0}}v^*_{\bk,m_{-1}(m_0-1)m_1}+\sqrt{m_{1}}\psi^{(0)}_{m_{1}}v^*_{\bk,m_{-1}m_0(m_1-1)} \right)
\\
\nonumber
&-&t\left(\sqrt{m_{-1}+1}\psi^{(0)*}_{m_{-1}}v^*_{\bk,(m_{-1}+1)m_0m_1} +\sqrt{m_{0}+1}\psi^{(0)*}_{m_{0}}v^*_{\bk,m_{-1}(m_0+1)m_1}+\sqrt{m_{1}+1}\psi^{(0)*}_{m_{1}}v^*_{\bk,m_{-1}m_0(m_1+1)} \right)
\\
\nonumber
&-&t\gamma_{\bk} \left(\sqrt{m_{-1}}V^+_{m_{-1}}A^{(0)}_{(m_{-1}-1)m_0m_1} + \sqrt{m_{0}}V^+_{m_{0}}A^{(0)}_{m_{-1}(m_0-1)m_1} + \sqrt{m_{1}}V^+_{m_{1}}A^{(0)}_{m_{-1}m_0(m_1-1)} \right)
\\
\nonumber
&-&t\gamma_{\bk}\left(\sqrt{m_{-1}+1}V^-_{m_{-1}}A^{(0)}_{(m_{-1}+1)m_0m_1} + \sqrt{m_{0}+1}V^-_{m_{0}}A^{(0)}_{m_{-1}(m_0+1)m_1}+\sqrt{m_{1}+1}V^-_{m_{1}}A^{(0)}_{m_{-1}m_0(m_1+1)}\right).
\end{eqnarray}
\end{widetext}
where
\begin{eqnarray}
V^+_{m_{\alpha}} &=& \sum_{\Bm'}\sqrt{m_{\alpha}'+1}
\\
\nonumber
&\times&\left(A_{\Bm'}^{*(0)}v^*_{\bk,(m_{\alpha}'+1)m_{\gamma}'m_{\delta}' }+A_{(m_{\alpha}'+1)m_{\gamma}'m_{\delta}' }^{(0)}u^*_{\bk,\Bm'}\right),
\\\nonumber
V^-_{m_{\alpha}} &=& \sum_{\Bm'}\sqrt{m_{\alpha}'}\times
\\
\nonumber
&\times&\left(A_{\Bm'}^{*(0)}v^*_{\bk,(m_{\alpha}'-1)m_{\gamma}'m_{\delta}' }+A_{(m_{\alpha}'-1)m_{\gamma}'m_{\delta}' }^{(0)}u^*_{\bk,\Bm'}\right).
\end{eqnarray}

Because of the mean field terms, the equations of motion for the $u_{\bk}$ are coupled to those of the $v_{\bk}$. These equations can be readily solved numerically and have the property that for every eigenvalue $\omega_{\bk}$, $-\omega_{\bk}$ is also an allowed eigenvalue. In the following section, we discuss in detail the lowest collective modes $\omega_{\bk}$ for repulsive and attractive $U_{2}$ across the entire superfluid-Mott insulator phase diagram. 

\section{Excitations}

\begin{figure*}[htp]
\begin{picture}(200, 90)
\put(-150, 100){\text{(a)}}
\put(25, 100){\text{(b)}}
\put(190, 100){\text{(c)}}
\put(-140, 0){\includegraphics[scale=0.45]{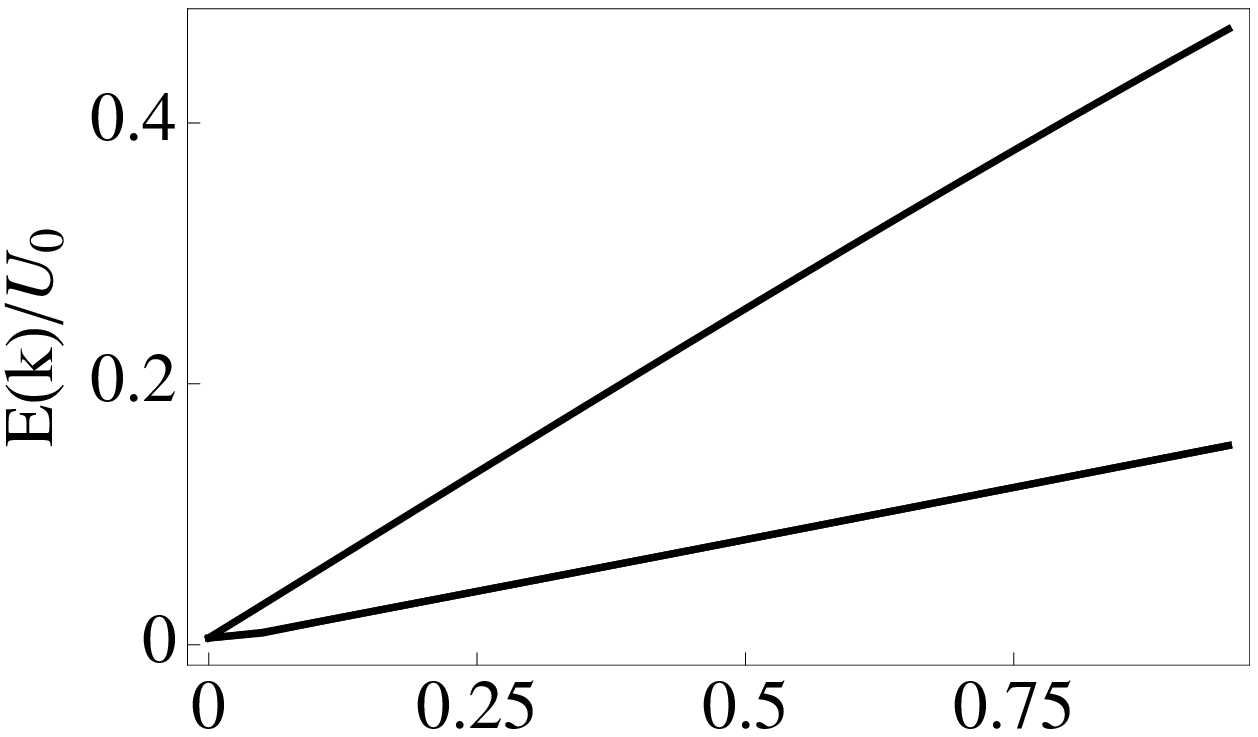}}
\put(30, -10){\includegraphics[scale=0.42]{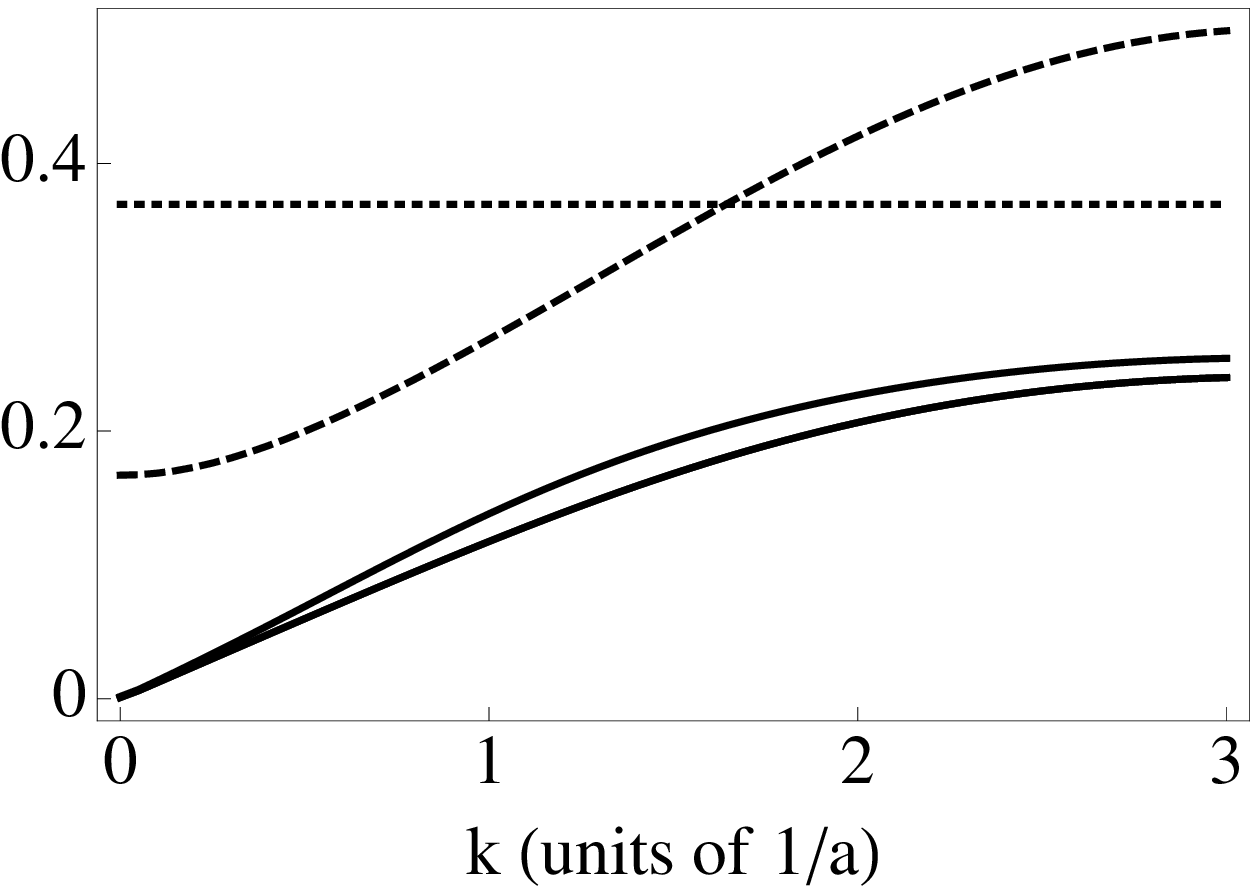}}
\put(190, 0){\includegraphics[scale=0.43]{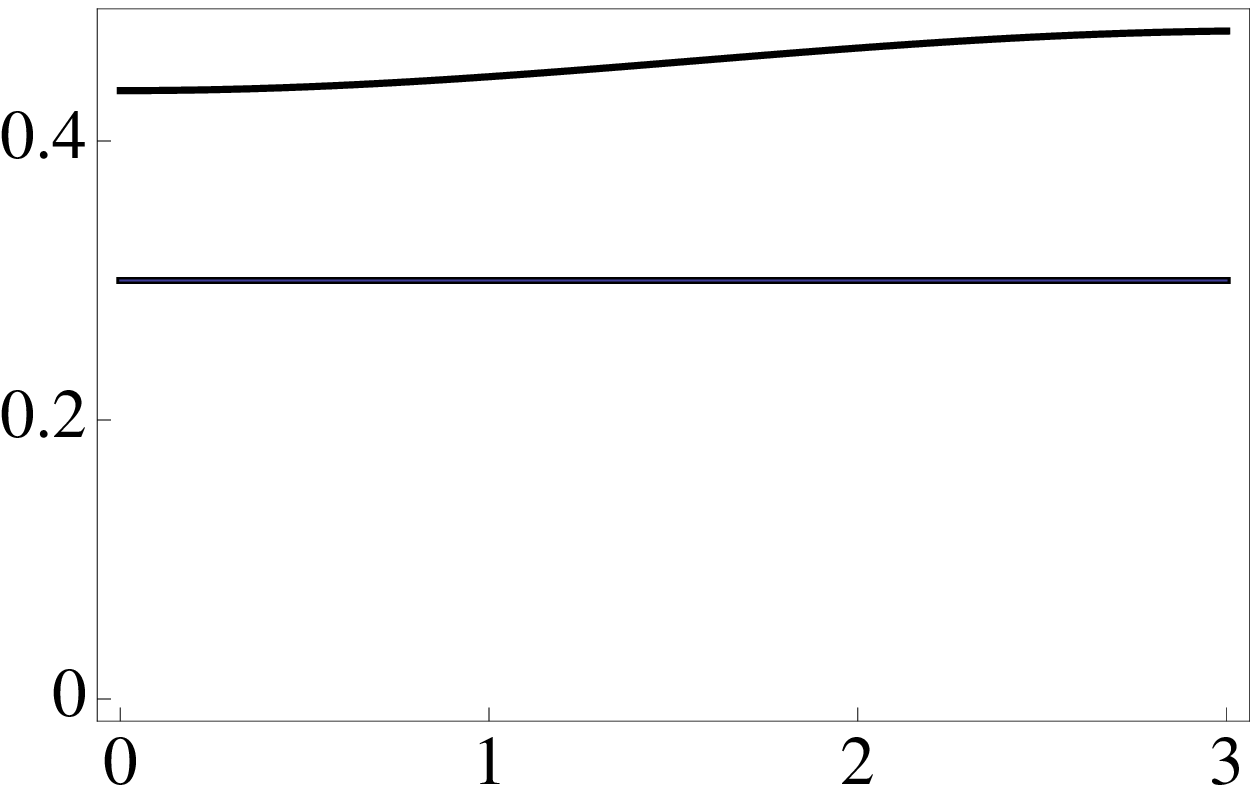}}
\end{picture}
\caption{\label{polarexfig} Evolution of the excitation spectrum from the nematic (polar) superfluid into the singlet phase. Throughout we fix $U_{2} = 0.1 U_{0}$ and vary $t$ and $\mu$ such that we access the deep superfluid (left),  superfluid-Mott phase boundary (center) and the deep Mott phase (right). The polar superfluid has two linearly dispersing modes corresponding to density and spin modes, originally derived in Refs.~\onlinecite{jason, machida}. As the superfluid-Mott phase boundary is approached, the sound speeds associated with the density and spin modes become identical. Additionally, there is a quadratically dispersing gapped, particle-hole mode and a non-dispersive mode. This non-dispersive mode corresponds to the breaking of a local singlet pair and has \textit{no} analog in the spinless case. This is therefore a new feature of the spin-$1$ Bose gas. Deep in the Mott insulator, the singlet pair breaking mode has the lowest energy and the particle-hole mode is pushed to higher energies.}
\end{figure*}

\subsection{Anti-ferromagnetic interactions}

We begin by discussing the case of anti-ferromagnetic interactions $U_{2} > 0$. In Fig.~\ref{polarexfig} we plot the evolution of the low lying excitation spectrum across the superfluid-singlet Mott phase diagram. Throughout, we fix $U_{2} = 0.1U_{0}$ and vary $t/U_{0}$ and $\mu/U_{0}$ with a density of $n \approx 2$, in order to access different regimes of the singlet Mott insulator-superfluid phase diagram. 

In the superfluid phase, we recover the three linearly dispersing modes; one density mode, and two degenerate spin modes, as originally described by the authors of Refs.~\onlinecite{jason, machida} for the continuum spin-$1$ Bose gas. The sound speeds associated with the density $c_{d}$ and spin modes $c_{s}$ are proportional to $\sqrt{U_{0}}$ and $\sqrt{U_{2}}$ respectively, and are typically vastly different in atoms such as $^{23}$Na (Ref. \onlinecite{jason}). As the Mott transition is approached however, density fluctuations are suppressed and consequently $c_{d}$ approaches $c_{s}$ monotonically. The sound speed associated with the spin mode remains relatively unaffected, as it is not directly related to the compressibility. As discussed in Sec.IIIA, the transition from the superfluid to the singlet-Mott insulator is first order, and is characterized by an abrupt jump in the superfluid order parameter. Thus at the Mott transition, the sound velocity also shows a discontinuous jump to zero (not shown). This is in contrast to the spinless case (or the ferromagnetic spin-$1$ gas or the nematic Mott-superfluid transition), where the sound speed either remains finite (at the Mott tip \cite{navez}) or goes to zero continuously, following the superfluid order parameter.  

\begin{figure}
\begin{picture}(200, 80)
\put(0, -10){\includegraphics[scale=0.45]{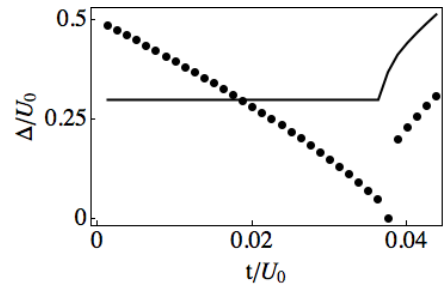}}
\end{picture}
\caption{\label{gaps} Evolution of the singlet gap (solid) and the particle-hole gap (points) across the superfluid $n=2$ singlet Mott insulator boundary for anti-ferromagnetic interactions. ($U_{2}/U_{0} = 0.1$). The singlet gap is completely independent of the hopping in the Mott insulator and and can be obtained by diagonalizing the on-site Hamiltonian. The particle-hole gap discontinuously goes to zero as the Mott-superfluid phase boundary is approached from the superfluid side, indicative of a first order transition. It approaches $0.5 U_{0}$ deep in the Mott insulator phase.}
\end{figure}

Near the Mott transition, but still in the superfluid phase, we additionally find two \textit{gapped} modes: a quadratically dispersing particle-hole mode and a non-dispersive mode which corresponds to the breaking of a singlet pair. The latter mode is absent in \textit{odd} integer Mott lobes, and is a new feature of the spin-$1$ gas, with no analog in the spinless case. 

In Fig.~\ref{gaps}, we plot the particle-hole gap $\Delta_{\text{ph}}$ 
and the singlet gap $\Delta_{\text{s}}$ 
across the superfluid-Mott phase boundary. The particle-hole gap evolves non-monotonically as the superfluid-insulator transition is crossed, going to zero at the phase boundary, signaling a quantum phase transition. For a second order quantum phase transition, the closing of the gap obeys the same power laws on either size of the transition scaling as $\Delta \sim |t-t_{c}|^{\nu}$ where $t_{c}$ is the critical point, and $\nu$ is the correlation length critical exponent. This is no longer true for a first order transition; as shown in Fig.~\ref{gaps}, the gap closes discontinuously when the transition is approached from the superfluid side, but continuously if approached from the Mott side. 

\begin{figure*}[htp]
\begin{picture}(200, 90)
\put(-150, 100){\text{(a)}}
\put(25, 100){\text{(b)}}
\put(190, 100){\text{(c)}}
\put(-140, 0){\includegraphics[scale=0.45]{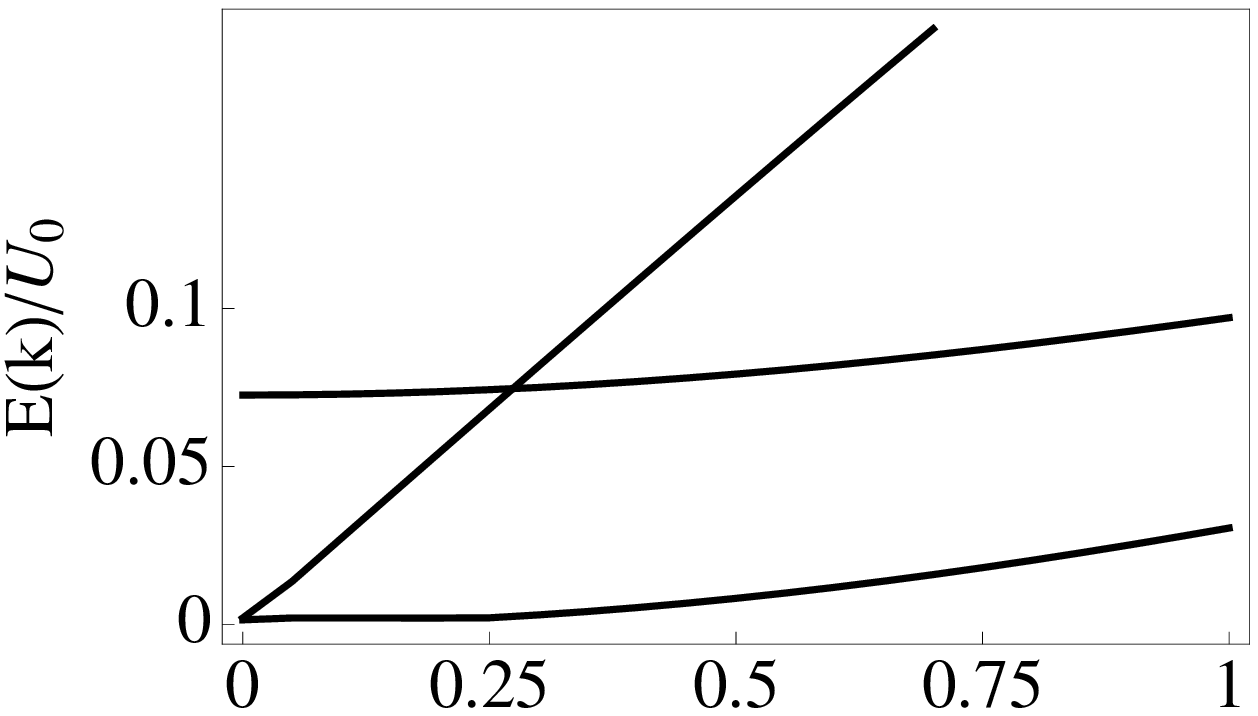}}
\put(30, -10){\includegraphics[scale=0.42]{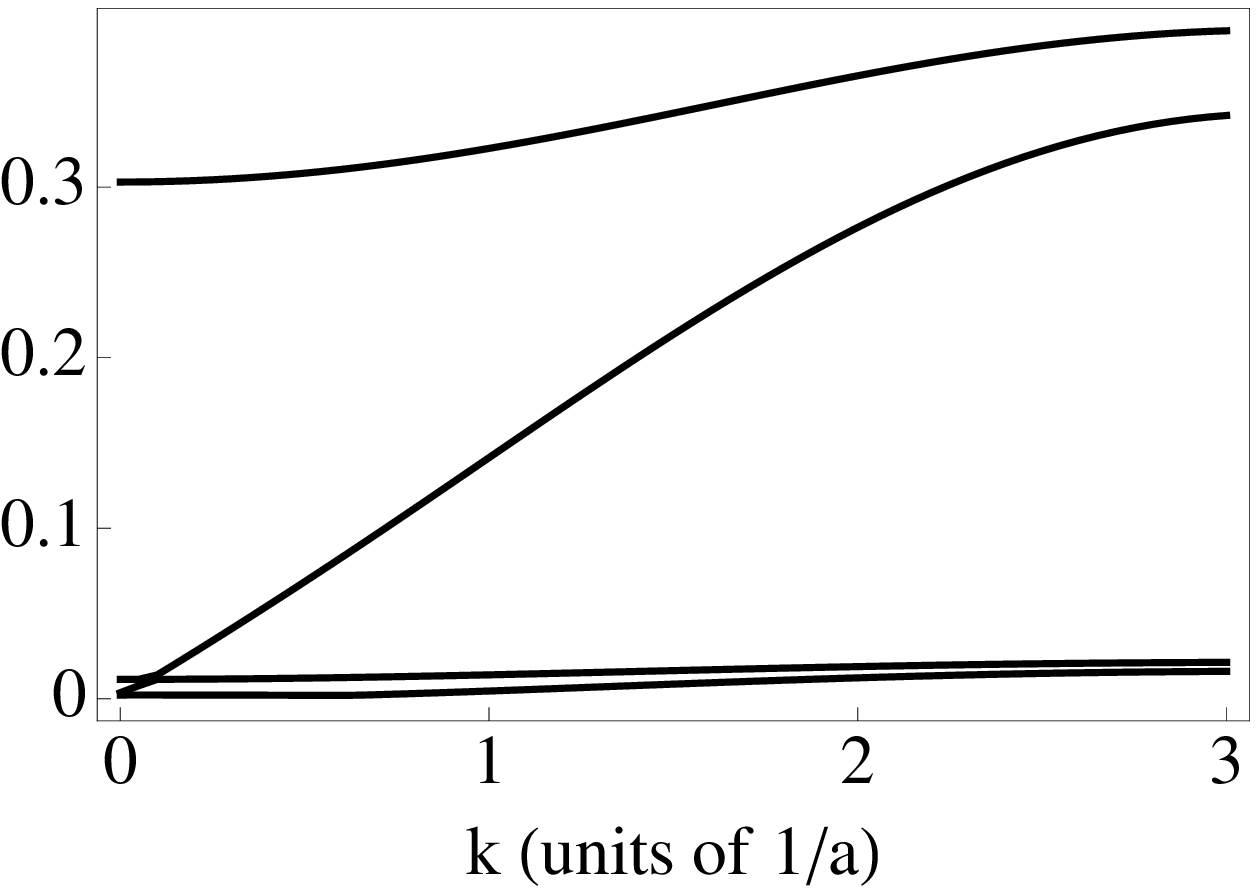}}
\put(190, 0){\includegraphics[scale=0.43]{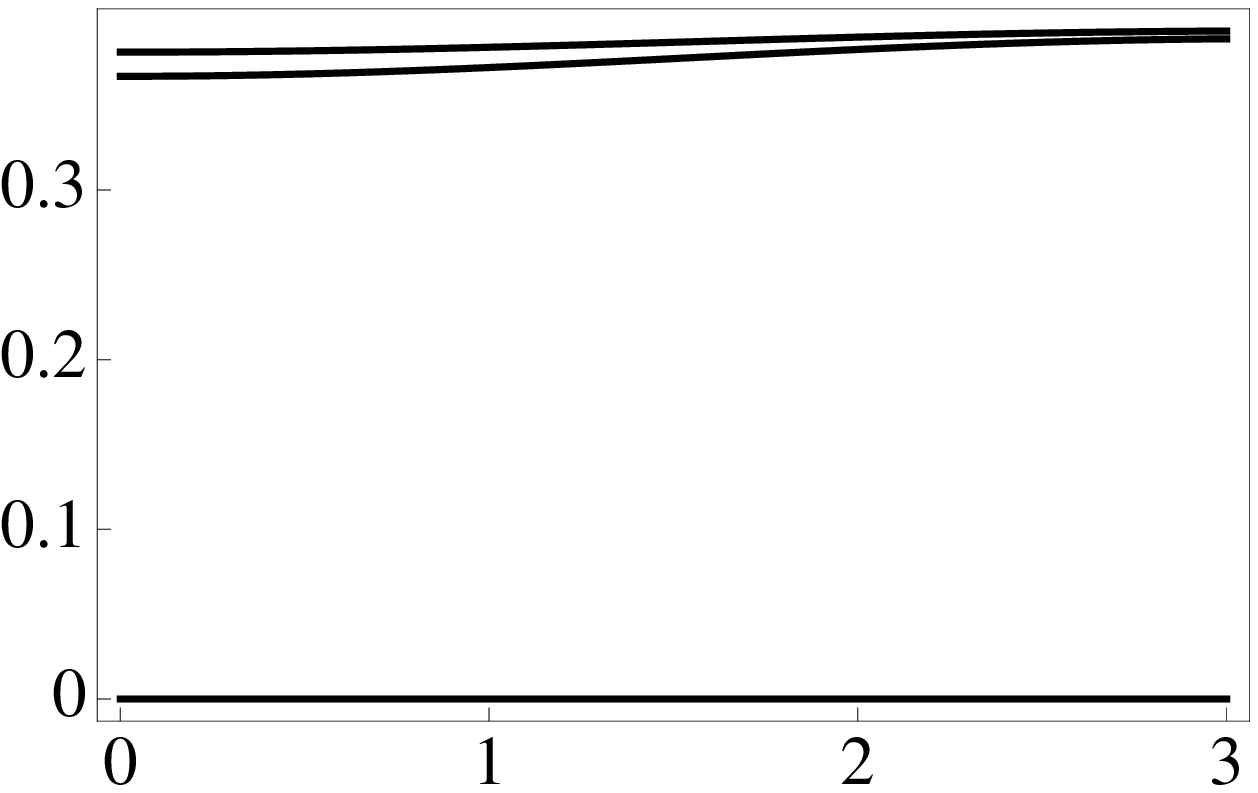}}
\end{picture}
\caption{\label{ferroexfig} Evolution of the excitation spectrum from the ferromagnetic superfluid into the Mott insulating phase. Throughout we fix $U_{2} = -0.1 U_{0}$ and vary $t$ and $\mu$ such that we access the deep superfluid (left),  superfluid-Mott phase boundary (center) and the deep Mott phase (right). The ferromagnetic superfluid has one linearly dispersing mode corresponding to density fluctuations and a quadratically dispersing gapless and a gapped spin mode, corresponding to fluctuations of the spin in the direction parallel and perpendicular to the easy axis. As the superfluid-Mott phase boundary is approached, the spin gap associated with the spin mode goes to zero. Additionally, there is a gapped, quadratically dispersing particle-hole like mode. The sound speed associated with the density mode vanishes at the transition and deep in the Mott insulator, there are two non-degenerate gapped modes corresponding to particle and hole-like excitations. The gapped spin mode is effectively non-dispersive in the Mott insulator as its effective mass is proportional to $t$ which is exponentially small.}
\end{figure*}

\begin{figure}
\begin{picture}(200, 90)
\put(-10, 0){\includegraphics[scale=0.45]{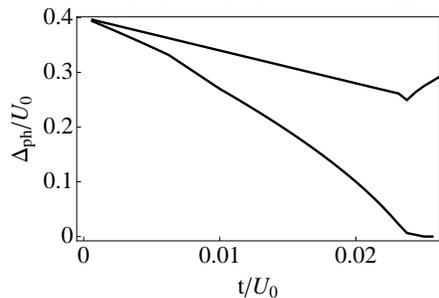}}
\end{picture}
\caption{\label{ferrogap} Evolution of the particle-hole gap across the superfluid- $n=2$ ferromagnetic Mott insulator boundary for $U_{2} = -0.1 U_{0}$.}
\end{figure}

Deep in the $n=2$ Mott insulator, the gap approaches $\Delta_{\text{ph}} \rightarrow U_{0}/2$ as $t \rightarrow 0$. By contrast, within the Gutzwiller approximation, the singlet gap is \textit{independent} of the hopping in the Mott insulator, and can be readily estimated by diagonalizating the onsite part of the Hamiltonian in Eq.~\ref{ham}. Within the Gutzwiller theory, the singlet and particle-hole modes do not couple to one another, and there is no hybridization gap as the particle-hole mode crosses the singlet mode. In reality, quantum fluctuations will couple the singlet and particle-hole modes and the singlet gap will depend on $t$. This goes beyond the Gutzwiller approach and has been studied by Imambekov \textit{et al.} \cite{imambekov} and Snoek and Zhou \cite{snoek}, and is not reproduced here. Their main result is that the singlet gap indeed varies with hopping and jumps to zero at the singlet-nematic Mott transition, which is not captured within our theory.  In general, the strong coupling theory of the even $n$ Mott lobes can be described by a constrained quantum rotor model~\cite{snoek}.  

In the nematic (odd $n$) Mott lobes, and the ferromagnetic Mott lobe for small $|U_{2}|/U_{0}$, the low energy description corresponds to a ferromagnetic, spin-$1$ bilinear-biquadratic $J-K$ spin Hamiltonian (see Eq. \ref{eqn:JK} below), originally derived by Imambekov \textit{et al.} \cite{imambekov} and Snoek and Zhou \cite{snoek}.  In the nematic Mott insulator, the low energy excitations are linearly dispersing nematic waves, whose spectrum was studied in detail by Imambekov \textit{et al.} \cite{imambekov}, and is not reproduced here. This theory is beyond the naive Gutzwiller approach we develop above, as the Gutzwiller theory does not contain any spin fluctuations in the Mott lobes. Below we study this biquadratic spin model on the ferromagnetic side, and compute the spin wave spectrum and discuss its evolution as a function of $t$.

\subsection{Ferromagnetic interactions}

We now turn to the case of ferromagnetic interactions $U_{2} <0$ which is shown in Fig.~\ref{ferroexfig}. Deep in the superfluid phase, we once again recover the excitation spectrum derived for the spin-$1$ Bose gas in Refs.~\onlinecite{machida, jason}. The low lying spectrum displays a single linearly dispersing mode corresponding to density excitations and two quadratically dispersing modes corresponding to the spin excitations about the ferromagnetic ground state. One of the modes has a free-particle spectrum which corresponds to spin waves while the other mode is gapped, and corresponds to ``quadrupolar" spin fluctuations \cite{jason}.

Deep in the Mott insulator, there are two non-degenerate low lying excitations which correspond to particle and hole like modes respectively. These modes are also present in the spinless case. At the Mott transition one of the gaps vanishes, signaling the phase transition into a superfluid and the other gap remains finite across the phase boundary as shown in Fig.~\ref{ferrogap}. Near the phase boundary but on the superfluid side, the effective mass of the spin mode decreases, as in a lattice, the effective mass is proportional to the hopping, which scales exponentially with the lattice depth. Furthermore, the spin gap associated with the quadrupolar spin mode vanishes at the transition. At the ferromagnetic transition the superfluid density vanishes, leading to a vanishing phonon velocity. 

Note that unlike the spinless gas however, the ferromagnetic Mott insulator has long range \textit{spin} order, and thus according to Goldstone's theorem, has a gapless mode corresponding to spin excitations. However this mode is not captured within the naive Gutzwiller approach as spin fluctuations in this theory are tied to the condensate order parameter, and therefore vanish at the Mott transition. To capture spin fluctuations in the insulating phase therefore, we augment the Gutzwiller theory with a Schwinger boson mean-field theory for the spin, which is described next. 

\subsubsection{Schwinger boson mean-field theory}
In the following subsection we determine the low lying excitations in the ferromagnetic Mott phase using the Schwinger boson mean field theory~\cite{assabook} (SBMFT).  We will focus on the $n=1$ Mott lobe for simplicity, but our results are also relevant to any value of $n$ with a change of the spin model parameters.  Working in the limit $U_0 \gg |U_2| \gg t$, we can apply standard perturbation theory in $t/(U_0+gU_2)$ (where $g$ is a small integer) which yields the spin-$1$ biquadratic model~\cite{imambekov,snoek}
\begin{equation}
H_{JK} = \sum_{\langle i,j\rangle} \left(-J \bS_i\cdot\bS_j -\frac{1}{2}K(\bS_i\cdot\bS_j)^2\right).
\label{eqn:JK}
\end{equation}
The parameters $J>0$ and $K/2>0$ are related to $t,U_0,U_2$ via Eq. 22 of Ref.~\onlinecite{imambekov}.
Focusing on the ferromagnetic ground state we can assume that the $K$ term is not sufficient to destroy the long range ferromagnetic order.  Thus, we can treat the biquadratic term at the mean field level, which yields $(\bS_i\cdot\bS_j)^2\rightarrow -\Phi_{ij}^2+2\Phi_{ij}\bS_i\cdot\bS_j$, where $\Phi_{ij}= \langle \bS_i\cdot\bS_j \rangle$, and therefore this decoupling essentially just renormalized the nearest neighbor ferromagnetic interaction to $J+K\Phi_{ij}$.  We generalize the spin symmetry from SU$(2)$ to SU$(N)$ and introduce Schwinger bosons through $\bS_i = b_{im}^{\dag} {\bf s}_{mn}b_{in}$, where ${\bf s}_{mn}$ are the generators of SU$(N)$ and $m,n=1,\dots,N$.  The $b$ operators must satisfy the constraint $\sum_{n}b_{in}^{\dag}b_{in}=NS$, and the SBMFT is exact in the limit $N\rightarrow \infty$.  We stress that the bosonic spinon operators $b_{in}$ are not the bosonic operators in the Bose-Hubbard model in Eq.~\ref{ham}.  Following standard SBMFT techniques we decouple the spin-spin interaction in the ferromagnetic channel through $\bS_i\cdot\bS_j = :\mathcal{F}_{ij}^{\dag}\mathcal{F}_{ij}:/N - S^2$, where $\mathcal{F}_{ij} = \sum_{n}b_{in}^{\dag}b_{jn}$, and $:\dots:$ denotes normal ordering.  Solving the SBMFT equations with $N=2$ and $S=1$ at zero temperature yields a ferromagnetic ground state~\cite{assabook} with a bosonic excitation spectrum given by
\begin{eqnarray}
\omega_{\bk} &=& z(J+K)-(J+K)\gamma_{\bk}
\\
&\underset{\bk\rightarrow 0}{\approx}& (J+K)|\bk|^2.
\end{eqnarray}
We recover the expected quadratically dispersing ferromagnetic spin waves with an effective mass $m^*= 1/(2[J+K])\sim U_0/t^2$ (in units of $\hbar=1$).  Therefore, we conclude that the correct excitation spectrum in the ferromagnetic Mott lobes have gapless Goldstone modes which disperse quadratically and cannot be captured within the Gutzwiller approach. We emphasize that Eq. ~\ref{eqn:JK} is only valid deep in the Mott insulator and \textit{not} near the transition where the truncation of basis states needed to arrive at this equation is no longer valid due to the vanishing of the particle-hole gap.

\section{Experimental Implications}

The continuum physics of the spin-$1$ Bose gas has been well studied experimentally, and the phase diagram is well established \cite{stenger, chang, stamper-kurn, stamperkurnreview, sadler, jacob}. By contrast, the phase diagram of the lattice spin-$1$ gas has received relatively little attention from the experimental community, despite the plethora of interesting phases and phase transitions present in this model. As we have shown here, the strongly correlated spin-$1$ superfluid and Mott regimes have many distinct features that are absent in the well studied spinless Bose Hubbard model \cite{greiner}. Indeed it will be extremely interesting to study the evolution of the nematic and singlet order parameters in a strongly correlated spin-$1$ gas with anti-ferromagnetic interactions, as in $^{23}$Na. Importantly, the evolution of these order parameters reveals a first order quantum phase transition near the superfluid-singlet Mott insulator, which has been confirmed numerically in $1$D. It will be very exciting if the predicted first order nature of the quantum phase transition can be explored experimentally.

The excitations in the spin-$1$ gas are also strikingly different from their spin-$0$ counterparts. For example, unlike the spinless Mott insulator, which is truly featureless, the ferromagnetic Mott insulator has quadratically dispersing spin waves, corresponding to long range spin order. The excitations in the weakly interacting superfluid limit of the spin-$1$ ferromagnetic gas were recently explored by Marti \textit{et al.} \cite{marti}, where a spin wave was externally imprinted on to the cloud and its coherent evolution was subsequently imaged. This method can also be applied in the Mott insulating regime to explore the spin wave spectrum in the ferromagnetic Mott lobe. Particle-hole like excitations couple to density fluctuations which are readily generated using modulation spectroscopy \cite{blochmod} or Bragg spectroscopy~\cite{ozeri}. For anti-ferromagnetic interactions, the low energy excitations are nematic waves, which are linearly dispersing in the superfluid and nematic Mott insulator. As is known from the theory of liquid crystals, the nematic tensor couples to the polarization of the incoming light beams, leading to optical birefringence, which can be used to probe nematic order and nematic waves \cite{carusotto}. 

\section{Conclusions and Outlook}

To conclude, in this paper we have presented a comprehensive mean-field description of the static and dynamic properties of the superfluid-Mott insulator transition in a spin-$1$ Bose gas. A key distinction in our work from previous works on this subject is our focus on the evolution of the important order parameters for the ferromagnetic and anti-ferromagnetic interactions, namely the spin in the ferromagnetic case, and the singlet and nematic order parameter in the anti-ferromagnetic case. 

Additionally, we have described the low lying excitation spectrum of the spin-$1$ gas across the entire phase diagram for positive and negative $U_{2}$. For $U_{2} >0$, we have discussed the evolution of the singlet and the particle-hole gap, which can be probed using modulation spectroscopy \cite{blochmod}. The singlet gap is a new feature of the spin-$1$ gas, and has no analog in the spinless case. We have discussed the limitations of the Gutzwiller approach in that it neglects the quantum fluctuations which couple the singlet and particle-hole modes. This would make the singlet gap vary as a function of tunneling, eventually going to zero at the singlet-superfluid transition. 

On the ferromagnetic side, we have discussed the evolution of the quadrupolar spin gap and the particle-hole like excitations across the phase diagram. Furthermore, we have pointed out a shortcoming of the Gutzwiller approach in describing spin fluctuations in the Mott insulator. Unlike the spinless Mott insulator, the ferromagnetic Mott insulator is not \textit{featureless} but rather is characterized by long range spin order. However, within this theory, spin fluctuations are tied to the condensate order parameter. Therefore, this theory accurately captures the spin modes in the condensate and reproduces the Bogoliubov spectrum at weak coupling. However, in the Mott insulator, where the condensate order parameter vanishes, spin fluctuations are frozen out, and as a result, the ferromagnetic Mott insulator does not have any spin fluctuations, in violation of Goldstone's theorem. To overcome this limitation, we have presented a Schwinger boson mean-field theory, which retains spin fluctuations in the Mott insulator and yields an additional free-particle like mode with an effective mass $m^{*}$ which varies like $U_0/t^2$ within the Mott insulating phase.  This mode is a new feature of the spin-$1$ gas, and can be probed using magnon interferometry \cite{marti}. 

Theoretically, the Gutzwiller approach developed here serves as a natural starting point for exploring more complicated Hamiltonians where the single particle physics involves a coupling between spin and kinetic degrees of freedom, such as the spin-orbit coupled Bose Hubbard model. The interplay between large spin and spin-orbit coupling can lead to simultaneous nematic, ferromagnetic orders with broken translational symmetry \cite{natusoc} or exotic spin models with novel ground states even at the classical level \cite{colesoc, radicsoc}. In the presence of single particle degeneracies such as those introduced by spin-orbit coupling, or artificial gauge fields, the absence of a bosonic Pauli principle severely limits exact numerical approaches, and only small system sizes can be accurately simulated numerically. Extending the Gutzwiller method to study the mean-field physics of these large spin, spin-orbit coupled models is therefore imperative \cite{arunnew, jedinprep}, and serves as a useful starting point for exploring the role of quantum fluctuations, the breakdown of mean-field theory and other strongly correlated effects, such as the fermionization of bosons in flat bands \cite{sedrakyan}. Importantly, this mean-field theory can be systematically generalized to incorporate fluctuation effects by solving the system exactly for small clusters, coupled by mean fields or by supplementing the Gutzwiller method with Schwinger bosons as done here to correctly capture low energy spin physics. 

\section{Acknowledgements}

We thank William Cole for useful discussions. This work is supported by the JQI-NSF-PFC and ARO-MURI.

\end{document}